\documentclass{emulateapj} 
\usepackage{amsmath}
\usepackage{amsfonts}
\usepackage{amssymb}
\usepackage{bm}
\usepackage{graphicx}% Include figure files
\usepackage{times}
\def\eps{\varepsilon}

\begin{document}

\title{A Kinetic Alfv\'en wave cascade subject to collisionless damping cannot reach electron scales in the solar wind at 1 AU}

\author{J.~J.~Podesta, J.~E.~Borovsky, and S.~P.~Gary}%\altaffilmark{1}}
\affil{Los Alamos National Laboratory, Los Alamos, NM 87545}
\email{jpodesta@solar.stanford.edu}

\begin{abstract}
Turbulence in the solar wind is believed to generate an energy cascade that is supported
primarily by Alfv\'en waves or Alfv\'enic fluctuations at MHD scales and by kinetic Alfv\'en 
waves (KAWs) at kinetic scales $k_\perp \rho_i\gtrsim 1$.  Linear Landau damping of KAWs 
increases with increasing wavenumber and at some point the damping becomes so strong
that the energy cascade is completely dissipated.  A model of the energy cascade 
process that includes the effects of linear collisionless damping of KAWs and the 
associated compounding of this damping throughout the cascade process is used to determine
the wavenumber where the energy cascade terminates.  
It is found that this wavenumber occurs approximately when 
$|\gamma/\omega|\simeq 0.25$, where $\omega(k)$ and $\gamma(k)$ are, respectively,
the real frequency and damping rate of KAWs and the ratio $\gamma/\omega$
is evaluated in the limit as the propagation
angle approaches 90 degrees relative to the direction of the mean magnetic field.
For plasma parameters typical of
high-speed solar wind streams at 1 AU, the model suggests that the KAW cascade in the 
solar wind is almost completely dissipated before reaching the wavenumber
$k_\perp \rho_i \simeq 25$.  Consequently, an energy cascade consisting solely of 
KAWs cannot reach scales on the order of the electron gyro-radius, $k_\perp \rho_e\sim 1$.  
This conclusion has important ramifications for 
the interpretation of solar wind magnetic field measurements.  
It implies that power-law spectra in the regime of electron scales
must be supported by wave modes other than the KAW.
\end{abstract}


\section{Introduction}

Recent papers by \citet{Schekochihin:2009}, \citet{HowesJGR:2008a}, and
\citet{Schekochihin:2008} have described a scenario
for turbulence in collisionless magnetized plasmas, such as the solar wind, that can be
briefly described as follows.  The turbulence at large scales, scales larger than the ion
inertial length $c/\omega_{pi}$ and the thermal ion gyro-radius $\rho_i$, consists of an 
energetically dominant Alfv\'en wave cascade that transfers energy from large to small 
scales.  At scales on the order of the proton gyro-radius, $k_\perp \rho_i \sim 1$, 
the wavevector spectrum of the turbulence is highly anisotropic with 
energy concentrated in wavevectors nearly perpendicular to the mean magnetic field 
$\bm B_0$ so that $k_\perp \gg k_\parallel$.  At
$k_\perp \rho_i \sim 1$, two things happen.  On the one hand, there may be some nonlinear 
effects that are not well understood.  On the other hand, a significant fraction of the 
energy in the Alfv\'en wave cascade excites a kinetic 
Alfv\'en wave cascade (KAW cascade) that carries the energy down to
scales on the order of the thermal electron gyro-radius 
where the turbulence is finally dissipated by collisionless Landau damping.
%Although whistler waves in the kinetic regime have not been included in this scenario,
%it is still possible that they may play a role at high wavenumbers.
%\smallskip

\citet{Sahraoui:2009} used the above scenario to interpret
spacecraft measurements of solar wind turbulence in the kinetic range of scales 
$k_\perp \rho_i\gtrsim 1$.  These authors used the linear KAW dispersion relation 
and damping rates to argue that the KAW cascade should reach electron scales before 
being strongly damped, noting that the relative damping rate $\gamma/\omega$ does not
become of order unity until $k_\perp \rho_e\sim 1$, where $\rho_e$ is the thermal 
electron gyroradius.  The purpose of the present paper is to
point out that this argument is incomplete
because it does not take into account the compounding of the damping throughout
the course of the cascade process.  KAWs are damped by collisionless Landau and 
transit-time damping
throughout the entire wavenumber range of their existence from 
$k_\perp \rho_i\simeq 1$ to $k_\perp \rho_i\gg 1$.  If the energy cascade process is  
thought of as taking place in a sequence of discrete steps, then there is dissipation at
each step in the sequence and the effects of damping are compounded with each step,
analogous to the way compound interest works.  When this compounding is taken into account
it is found that for typical solar wind plasma parameters near 1 AU the KAW cascade is 
dissipated before reaching electron scales.  This conclusion has important ramifications 
for the interpretation of solar wind power spectra in the kinetic regime. 
%\smallskip

In astrophysical plasmas, the effects of collisionless damping on the turbulence spectrum 
have been studied primarily using quasilinear theory \citep[see, for example,][]{Melrose:1994}
and phenomenological models in which the 
energy cascade process is modeled as a diffusion process in wavenumber space
\citep[for example,][]{Li:2001,Stawicki:2001,Cranmer:2003,Jiang:2009,Matthaeus_PRE:2009}.  
Other approaches include weak turbulence theory 
\citep{Yoon:2006,Yoon:2007,Galtier:2006,Chandran:2008a} and 
gyro-kinetic theory \citep{HowesPRL:2008b, Schekochihin:2009}.
Here we adopt a somewhat simpler approach that, like the diffusion models,
is based on an equation expressing the conservation of energy in wavenumber
space.  Generally speaking, the objective of any of these models is to describe 
the essential physics as accurately and economically as possible.  
%\smallskip

In this study, the effects of damping on the KAW cascade are modeled using two 
complimentary approaches.  
%In both approaches the linear damping rates of KAWs are obtained from the hot plasma dispersion relation \citep{Stix:1992}.
The first approach is heuristic and shows how this damping 
takes place through a sequence of steps in wavenumber space whereby energy is damped at 
each step before being transferred to higher wavenumbers.  The second approach is based on
an equation for the conservation of energy in wavenumber space and two different 
expressions for the energy cascade rate that are similar to Kolmogorov's relation 
$\eps = kE(k)/\tau$.  This approach has much in common 
with the cascade model developed by \citet{HowesJGR:2008a} and
allows both the energy cascade rate $\eps$ and the energy spectrum $E(k)$ to be computed as
functions of wavenumber throughout the inertial range and dissipation range.  In this study,
the main objective is to determine the point in wavenumber space where the KAW cascade 
terminates and, therefore, the energy cascade rate $\eps$ is the quantity of 
primary interest.   Assuming that the propagation angles
of the waves are nearly perpendicular to $\bm B_0$, analytic solutions for the
energy cascade rate $\eps$ may be derived that are convenient for purposes of
analysis and prediction.  These analytic solutions are new 
and are presented here for the first time.
%\smallskip

One of the principal assumptions of this work is that 
%nonlinear interactions
%are sufficiently weak in the kinetic regime that the turbulence at those scales can be 
%treated as an ensemble of weakly interacting linear waves for which
the damping rates of linear wave theory are applicable at kinetic scales, even though 
nonlinear interactions are still strong.  
%If the nonlinear interactions
%are not sufficiently weak, then the fluctuations at those scales will not behave like
%linear waves and this approximation will break down. 
Although this same assumption
has been made before by many investigators \citep{Leamon_Smith:1998,Quataert:1998,
Gary:1999,Leamon:1999c,Quataert:1999,Li:2001,Stawicki:2001,Cranmer:2003,HowesJGR:2008a}, 
there is no well established criteria for its validity.
This is, perhaps, the greatest source of uncertainty for the theory presented here.
It is also easy to envision  a scenario in which turbulence at MHD scales is
dissipated via collisionless magnetic reconnection involving structures (reconnection sites) 
that span lengthscales from the ion inertial length to the electron inertial length
\citep[][Chapter 3]{Birn_Priest:2007}.
To some extent, this  
%does not depend solely on collisionless wave damping to dissipate the 
%energy cascade and, therefore, it is 
is a possible alternative to the wave cascade and damping scenario 
described by \citet{Schekochihin:2009} and others.

For the moment, adopting the KAW cascade and linear damping scenario and assuming that the 
wave damping rates of the linear Vlasov-Maxwell theory are valid in the kinetic regime, then
for typical solar wind conditions at 1 AU the models derived here predict that the 
KAW cascade terminates at wavenumbers less than or equal to $k_\perp \rho_i\simeq 25$
which implies that the KAW cascade cannot reach electron scales in the solar wind 
at 1 AU.  These results agree with and are supported by those obtained previously 
using the cascade model of \citet{HowesJGR:2008a} which show a transition to an
exponentially decaying magnetic energy spectrum that decays rapidly at around the 
same wavenumber.  On the contrary, 
the magnetic energy spectra in the gyro-kinetic simulations of \citet{HowesPRL:2008b} 
do not appear to show any deviations from power-law behavior in the range 
$1\lesssim k_\perp \rho_i \lesssim 8$ as there should be if kinetic damping is
having a significant effect.  However, because these scales are near
the smallest spatial scales in the simulation where an artificial electron
hypercollisionality takes effect, the simulation may not   
accurately describe the physics when $k_\perp \rho_i \sim 8$.  
%\medskip

One note on terminology.
The term ``energy cascade'' or ``cascade,'' for short, refers to a
nonlinear energy transfer process that transfers energy from large to
small scales or vice versa.  This term usually refers to the inertial range
where the effects of dissipation are negligible.
Here, this term is also used to describe the nonlinear energy transfer in the
dissipation range.  This slight abuse of notation should cause no confusion.

%It has been suggested that at kinetic scales the energy cascade in the solar wind is 
%dominated by a cascade of KAWs.  Assuming  that this is true, then it is of interest to
%ask: What is the range of wavenumbers over which the KAW cascade is damped and what 
%fraction of the energy in the cascade will reach electron scales (the electron inertial 
%length and the electron gyro-radius)?  
%The answer, of course, depends on the plasma parameters in the problem.
%For plasma parameters typical of high-speed solar wind streams at 1 AU, we shall
%argue that the KAW cascade is damped continuously by linear Landau damping as the energy 
%cascades from the spectral break, $k_\perp \rho_i \simeq 1$, to around
%$k_\perp \rho_i \simeq 30$ and that only a small fraction ($<1\%$) of the energy in the
%cascade, if any, is capable of reaching electron scales.  
%\medskip

\section{Simple heuristic model}

%\indent
\indent
We begin with some general considerations that apply to the entire paper.
Consider a homogeneous proton-electron plasma with a constant uniform 
background magnetic field.  The equilibrium state is charge neutral, free of macroscopic 
electric fields, free of electric currents, and characterized by isotropic Maxwell 
distribution functions for both particle species.  It is assumed that the wave 
amplitudes at kinetic scales are small enough that linear wave theory adequately 
describes the collisionless damping process.  For a given wavenumber $\bm k$ (real-valued),
the real and imaginary parts of the wave 
frequency $\omega+i\gamma$ may be computed numerically using the hot plasma dispersion 
relation \citep[chap 10]{Stix:1992}.  For $k_\perp \gg k_\parallel$, the KAW is uniquely 
identified by means of its asymptotic dispersion relation $\omega = k_\parallel v_A$ 
in the limit as $k_\perp \rightarrow 0$ with $k_\perp/k_\parallel$ or the
angle of propagation held constant.
%\smallskip

In general, the damping is minimized for all
wavenumbers in the range from $k_\perp \rho_i \simeq 1$ to $k_\perp \rho_i \simeq 40$ when the
angle of propagation $\theta$ is close to $90^\circ$, that is, near perpendicular to $\bm B_0$.
It is important to note that for all angles in a sufficiently small neighborhood of $90^\circ$, 
the ratio of the damping rate to the real frequency, $\gamma/\omega$,
is approximately independent of the angle $\theta$ so that all propagation
angles in this range yield the same damping per wave period.  In practice,
this means that the damping per wave period cannot be reduced further by going from 
$89.9^\circ$ to $89.99^\circ$, for example.  For simplicity, it will be assumed in this
section that $\theta$ is close, but not equal to $90^\circ$.  This range of near-perpendicular 
angles is physically relevant because as a consequence of three-wave interactions the 
energy cascade process creates a spectrum in which the inequality 
$k_\perp \gg k_\parallel$ is increasingly well satisfied as the cascade progresses
to higher wavenumbers.  
%\medskip

Investigation of the damping of the KAW cascade begins with the plot in Figure-1
%-----------------------------------------------------------------------------------------------
\begin{figure}
\begin{center}
\includegraphics[width=3in]{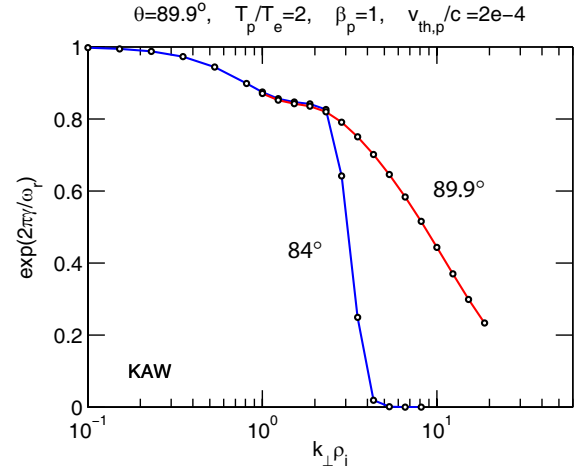}% 
\caption{\label{fig1}%
     Damping rate of the wave amplitude in one wave period $\alpha(k)=\exp[2\pi\gamma(k)/\omega(k)]$.
The propagation angle $\theta=84^\circ$ is shown in blue and $\theta=89.9^\circ$ is shown in red.
}
\end{center}
\end{figure}
%-----------------------------------------------------------------------------------------------
which shows the attenuation of the wave amplitude in one wave period, that is,
$\exp(2\pi\gamma/\omega)$ versus $k_\perp$ for a fixed angle of propagation close
to $90^\circ$. As just mentioned, the curve for $89.9^\circ$ in Figure-1 applies
for all angles sufficiently close to $90^\circ$.   Now recall the critical balance 
hypothesis of \citet{Goldreich_Sridhar:1995,Goldreich_Sridhar:1997} which states that 
in the {\it inertial range} the cascade time is equal to the Alfv\'en wave period.  
Using this as a guide, we assume that the cascade time in the {\it kinetic regime}
is equal to the wave period of the KAWs.  Consider the sequence of wavenumbers
$k_n=2^nk_0$, where $k_0 \rho_i=1$ and $n=0, 1, 2,\ldots$.
In one cascade time, one wave period, the energy at scale $k_n$ cascades to scale $k_{n+1}$
but in the process that energy is damped by a factor $\alpha_n^2$, where 
$\alpha_n=\exp[2\pi\gamma(k_n)/\omega(k_n)]$ is the attenuation of the wave amplitude 
in one wave period.  If $\eps(k_n)$ is the energy cascade rate at wavenumber $k_n$,
that is, the rate at which energy reaches scale $k_n$, then the rate at which energy reaches 
the next scale $k_{n+1}$ is 
\begin{equation}
\eps_{n+1}=\alpha_n^2 \eps_n,
%\label{diss}
\end{equation}
where $\eps_n=\eps(k_n)$.
This is the change in the energy cascade rate caused by linear damping after just one step 
in the cascade process.  Starting from $n=0$, the energy cascade rate after $n$ steps is equal to 
\begin{equation}
\eps_{n}=(\alpha_0\alpha_1\alpha_2\cdots\alpha_{n-1})^2 \eps_0.
\label{eps}
\end{equation}
This describes how the energy of the waves is damped as it cascades through the kinetic 
range of scales.  To compute the wavenumber dependence of  $\eps(k)$ for the KAW cascade 
it is only necessary to compute the relative damping rate $\gamma/\omega$ from the hot 
plasma dispersion relation and then use equation (\ref{eps}).
%\smallskip

%%%%%%%%%%%%%%%%%%%%%%%%%%%%%%%%%%%%%%%%%%%%%%%%%%%%%%%%%%%%%%%%%%%%%%%%%%%%%%%%%%%%%%
\begin{deluxetable}{cccr}
%\rotate
\tablewidth{0pt}
%\tabletypesize{\small}
\tablecaption{Cascade rates with compounding taken into account}
\tablehead{
  \colhead{$n$} &  \colhead{$k_\perp \rho_i$} &  \colhead{$\alpha_{n}$} &  \colhead{$\eps_n/\eps_0$} 
}
\startdata
0  & 0.5 &  $0.95$  &   100\%   \\
1  &  1 &  0.87 &   90\%     \\
2  &  2 &  0.84 &   68\%     \\
3  &  4 &  0.73 &   48\%     \\
4  &  8 &  0.52 &   25\%     \\
5  & 16 &  $<0.4$   &     7\%     \\
6  & 32 &   --- &   $< 1$\%     
\enddata
%\tablenotetext{a}{Spacecraft: STA $=$ Stereo A, STB $=$ Stereo B}
%\tablecomments{Data are from figure 1 and equation (\ref{eps}).}
\end{deluxetable}
%%%%%%%%%%%%%%%%%%%%%%%%%%%%%%%%%%%%%%%%%%%%%%%%%%%%%%%%%%%%%%%%%%%%%%%%%%%%%%%%%%%%%%
By way of illustration, consider the damping rates for $89.9^\circ$ shown in Figure-1.  
For each wavenumber $k_n=2^nk_0$ it is straightforward to read off the values of $\alpha_n$ 
from the plot in Figure-1 and then compute the cascade rate $\eps_n$ from equation 
(\ref{eps}).  This yields the results in Table-1 
which show that the energy cascade rate is reduced to
less than 1\% of its original value by the time the cascade reaches $k_\perp \rho_i=32$.
The plasma parameters used to compute the damping rates in Figure-1 are
typical of high-speed solar wind streams at 1 AU.  For these parameters, the electron 
gyro-radius occurs when $k_\perp \rho_e\simeq 1$ or, equivalently, $k_\perp \rho_i\simeq 60$, 
and the electron inertial scale occurs when $k_\perp c/\omega_{pe}\simeq 1$ or, equivalently, 
$k_\perp \rho_i\simeq 30$. 
This simple calculation indicates that the KAW cascade is likely to be dissipated before 
reaching electron scales.  As we show below, a more detailed model yields results that 
are in good agreement with those in Table-1.
%\medskip

\section{Model based on conservation of energy}

%\indent
\indent
%The model is first developed for hydrodynamic turbulence and the results are compared to 
%experimental measurements in the near dissipation range, that is, the range of scales
%greater than the Kolmogorov scale $\eta$.  The impatient reader may skip the following subsection
%without loss of continuity.  The application to hydrodynamic turbulence illustrates the 
%approach and allows an assessment of the results compared to experimental measurements.
The energy spectrum $E(k)$ is defined such that $E(k)\, dk$ is the energy per unit mass
contained in the fluctuations for all wavenumbers between $k$ and $k+dk$.  
The total energy contained in the entire spectrum is
\begin{equation}
E_{tot}=\int_0^\infty E(k)\, dk.
%\label{diss}
\end{equation}
The cascade rate $\eps(k)$ is the average energy per unit mass 
that passes through wavenumber $k$ per unit time.  The energy flows from low to high 
wavenumbers.  Consider a small interval from $k$ to $k+dk$ in wavenumber space as 
illustrated in Figure 2.
%-----------------------------------------------------------------------------------------------
\begin{figure}
\begin{center}
\includegraphics[width=2.9in]{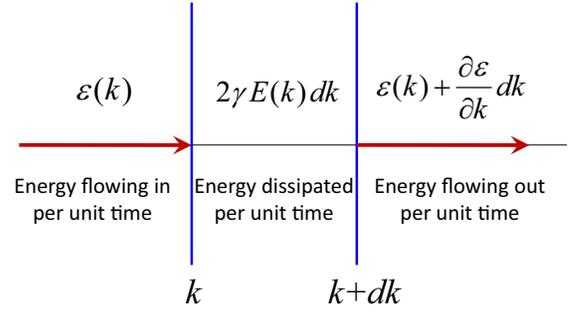}% 
\caption{\label{fig2}%
     For an infinitesimal interval from $k$ to $k+dk$, the conservation of energy in 
wavenumber space requires that the energy flowing
into the layer at wavenumber $k$ minus the energy dissipated in the layer equals the
energy flowing out at $k+dk$.  This yields equation (\ref{CE}).
}
\end{center}
\end{figure}
%-----------------------------------------------------------------------------------------------
Energy is flowing into the boundary $k$ from the left and out of the boundary $k+dk$ 
from the right.  Conservation of energy requires that the 
energy flowing out per unit time is equal to the energy flowing in per unit time
minus the rate at which energy is dissipated in the layer.  Hence,
\begin{equation}
\eps(k+dk)-\eps(k) = 2\gamma E(k)\, dk,
\label{cons}
\end{equation}
where $\gamma<0$ is the linear damping rate obtained from the KAW dispersion relation.
This yields
\begin{equation}
\frac{d\eps}{dk} = 2\gamma E(k)
\label{CE}
\end{equation}
which expresses the conservation of energy in the cascade process. Note that
if $\gamma=0$, then $\eps=$ {\it const} and the energy cascade rate 
is independent of $k$.  
%\smallskip

Equation (\ref{CE}) gives one relation between $\eps$ and $E$.  A second is
Kolmogorov's relation 
\begin{equation}
\eps(k) = \frac{kE(k)}{\tau},
\label{KR}
\end{equation}
where $\tau$ is the energy cascade time.  Komlogorov's relation assumes there is 
no dissipation of energy during the cascade process so that {\it all} the
energy at scale $k$ cascades to higher wavenumbers.  When dissipation is present,
a fraction $\alpha^2$ of the total energy at scale $k$ cascades
to higher wavenumbers and a fraction $1-\alpha^2$ is dissipated.  
The energy that is dissipated does not cascade to higher wavenumbers. 
In this case, the relation (\ref{KR}) is replaced by 
\begin{equation}
\eps(k) = \frac{\alpha^2 kE(k)}{\tau},
\label{KR2}
\end{equation}
where $\alpha^2(k)$ is the energy attenuation factor in time $\tau$.  
%\smallskip

It is clear
from physical considerations  that in the dissipation range the energy cascade rate 
$\eps$ must depend on the dissipation rate.  Hence, Kolmogorov's relation 
(\ref{KR}) that was initially postulated for the inertial range needs to be modified.  
What is not clear is what the precise functional form should be.
Justification for the form (\ref{KR2}) can be obtained by considering a simple shell model. 
Let $k_n=2^nk_0$ denote a sequence of shells in wavenumber space, where $n$ is an integer 
and $k_0$ is an arbitrary wavenumber.  The energy contained in the interval between $k_n$ 
and $k_{n+1}$ is $k_nE(k_n)$.  
In the absence of dissipation, this energy cascades to scale $k_{n+1}$ in one
cascade time $\tau(k_n)$ so that the energy cascade rate is
\begin{equation}
\eps_n = \frac{k_nE(k_n)}{\tau_n}.
%\label{KR}
\end{equation}
When there is no dissipation, $\eps_n = $ {\it const} and the energy cascade rate is 
independent of scale.
When dissipation is present, the energy at scale $k_n$ is partly dissipated before it can
cascade to the next level.  If the energy is attenuated by a factor $\alpha^2$ in
time $\tau_n$, then the energy cascade rate becomes
\begin{equation}
\eps_n = \frac{\alpha^2 k_nE(k_n)}{\tau_n},
%\label{KR}
\end{equation}
where $\alpha$ may depend on $k_n$. This is another way to derive equation 
(\ref{KR2}).  
%\smallskip

In general, equations (\ref{KR}) and (\ref{KR2}) should also contain a constant 
coefficient $A$ which is dimensionless.  In hydrodynamic turbulence it follows from
dimensional analysis that this coefficient is of order unity.  In plasma turbulence 
there are other parameters in the problem which may cause this coefficient to differ 
from unity. For example, in the inertial range, one such parameter is the normalized 
cross-helicity $\sigma_c$.  Thus, equation (\ref{KR2}) takes the final form
\begin{equation}
\eps(k) = A\frac{\alpha^2 kE(k)}{\tau},
\label{KR3}
\end{equation}
where in many cases of practical interest the dimensionless constant $A$ is of
order unity.
%\smallskip

\subsection{Complete model}
 
Assuming that 
critical balance holds for the KAW cascade, then the energy cascade time is equal to the 
wave period, $\tau=2\pi/\omega$, and the energy cascade rate (\ref{KR3}) has the form
\begin{equation}
\eps(k) = \frac{\omega (k)}{2\pi}A \exp(4\pi \gamma/\omega)kE(k),
\label{reln2}
\end{equation}
where $\alpha =\exp(2\pi \gamma/\omega)$ is the attenuation factor for the wave amplitudes
in one wave period and $\alpha^2 = \exp(4\pi \gamma/\omega)$ is the attenuation factor for 
the energy.  When equation (\ref{reln2}) is substituted into the conservation
law (\ref{CE}), it follows that
\begin{equation}
\frac{k}{\eps}\frac{d\eps}{dk} = \frac{1}{A} \frac{4\pi\gamma}{\omega}\exp(-4\pi \gamma/\omega) 
\label{CE2}
\end{equation}
or, equivalently,
\begin{equation}
\frac{d(\log\eps)}{d(\log k)} = \frac{1}{A} \frac{4\pi\gamma}{\omega}\exp(-4\pi \gamma/\omega).
\label{CE2log}
\end{equation}
When the exponential factor on the right-hand side is omitted, this equation is almost
identical to equation (8) in \citet{HowesJGR:2008a} with the source term $S$ in 
\citet{HowesJGR:2008a} set equal to zero.  The one minor difference is that the damping
rates $\gamma$ and $\omega$ in (\ref{CE2log}) are obtained from the hot plasma dispersion 
relation \citep{Stix:1992} whereas 
\citet{HowesJGR:2008a} employ the damping rates obtained from gyro-kinetic theory.

Because $\omega>0$ and $\gamma<0$, the exponential factor in (\ref{CE2log}) causes 
$\eps(k)$ to decrease more rapidly than it would if the exponential factor were omitted.  
When $|4\pi \gamma/\omega|>1$, the exponential drives $\eps$ toward zero at a
faster than exponential rate as can be seen in the solutions presented below.  At this point, 
the dissipation becomes so effective that the energy cascade is abruptly terminated.  
However, even if the exponential factor is omitted from equation (\ref{CE2log}) which is 
equivalent to using equation (\ref{KR}) in place of (\ref{KR2}), the wavenumber 
where the cascade rate $\eps$ goes to zero is still roughly of the same order of magnitude.  
Therefore, the presence of this exponential factor is not crucial for the conclusions
reached in this study.  
%The solutions given in the next section
%show that when the exponential factor is omitted from equation (\ref{CE2log}) the 
%solution is much more sensitive to variations in the coefficient $A$.  
%\smallskip

If $\gamma(k)$ and $\omega(k)$ are known from the hot plasma dispersion relation,
then equation (\ref{CE2log}) may be solved to find how $\eps$ depends on $k$.  Once the
solution for $\eps(k)$ is known, the energy spectrum $E(k)$ is obtained from equation 
(\ref{reln2}).  In general, the ratio $\gamma/\omega$ depends on the angle of 
propagation of the waves which needs to be taken into account.  However, when $k_\perp\gg 
k_\parallel$ or, equivalently, when the angle of propagation is sufficiently close to $\pi/2$,
then the ratio $\gamma/\omega$ becomes approximately independent of angle.
This property is used to derive an approximate analytic solution below.
In cases where the angle is not sufficiently close to $\pi/2$ the angle dependence
may be taken into account as follows.  Assume that 
the energy in the wavevector spectrum is concentrated near the critical balance curve 
so that the angle of propagation $\theta$ at a given scale $k_\perp =k$ is 
determined by the critical balance relation
\begin{equation}
\cot(\theta) = \frac{k_\parallel}{k_\perp} = \frac{\delta v_\perp}{v_{\parallel,\rm ph}}, 
\label{CB}
\end{equation}
where 
%$\delta b_\perp =[k_\perp E(k_\perp)]^{1/2}$ is the rms amplitude of the magnetic 
%field fluctuation at scale $k_\perp$ and 
$\delta v_\perp$ is the electron velocity perturbation of the wave in the plane 
perpendicular to $\bm B_0$ and $v_{\parallel,\rm ph}$ is the parallel phase speed of the KAWs.  
Note that the parallel phase speed can be much larger than the Alfv\'en speed in the
kinetic regime and this needs to be included in the critical balance relation (\ref{CB}).

According to the critical balance hypothesis, the longitudinal
crossing time of two wavepackets $\tau_\parallel=\lambda_\parallel/v_{\parallel,\rm g}$
is equal to the nonlinear eddy turnover time $\tau_\perp=\lambda_\perp/\delta v_\perp$,
where $v_{\parallel,\rm g}$ is the parallel group speed.  The parallel phase and group
velocities are approximately equal as can be seen from the approximate dispersion relation
$\omega \propto k_\parallel v_A \sqrt{1+(k_\perp \rho_i)^2}$ \citep[see, for example,][]
{Hollweg:1999,Cranmer:2003}.
%$v_A$ is the Alfv\'en speed based on the mean magnetic field $B_0$ at large scales or the total 
%rms value $B_{\rm rms}$ if there is no large scale magnetic field.  
In addition to equation (\ref{CB}), the self consistent calculation of the propagation 
angle $\theta$ requires the relation 
\begin{equation}
k E(k) \simeq (\delta b_\perp)^2, 
%k_\perp E(k_\perp) = (\delta v_\perp)^2+(\delta b_\perp)^2, 
\label{CB1}
\end{equation}
where the magnetic field perturbation $\delta b_\perp$ is measured in velocity units.
For the plasma parameters of interest here, the energy density of KAWs is dominated by 
magnetic and thermal energy (i.e., density fluctuations) with an approximate equipartition 
between magnetic and thermal energy as discussed, for example, by \citet{Terry:2001}.  
These two contributions have been lumped together into a single term in (\ref{CB1}). 
Critical balance tightly couples the magnitude of the energy spectrum to the propagation
angle through equations (\ref{CB}) and (\ref{CB1}).  
For KAWs, the ratio $\delta v_\perp/\delta b_\perp$ may be determined from
the hot plasma dispersion relation using the relations 
$\bm J_e = -i\omega \eps_0 \bm \chi_e\cdot \bm E=-en_0\bm v_e$, where $\bm J_e$ is the 
electron current density and $\bm \chi_e$ is the electron susceptibility \citep{Stix:1992}.
This completes the specification of the model which consists of equations 
(\ref{reln2}), (\ref{CE2log}), (\ref{CB}), and (\ref{CB1}).
%\medskip

\subsection{Analytic model}

The above model can be significantly simplified when $k_\perp\gg k_\parallel$ so that 
the ratio $\gamma/\omega$ becomes approximately independent of the propagation angle
$\theta$.  Simulations of incompressible MHD turbulence have shown
that the inequality $k_\perp\gg k_\parallel$ is usually well satisfied at the smallest 
inertial range scales and, therefore, the assumption $k_\perp\gg k_\parallel$ is
justified in the kinetic regime.  It is useful to adopt the approximate expressions for 
the ratio $\gamma/\omega$ derived by \citet{Howes:2006} using gyrokinetic theory.
Using equations (62) and (63) in \citet{Howes:2006} one obtains
\begin{equation}
\frac{\gamma}{\omega} \simeq -a k_\perp \rho_i,
\label{ratio}
\end{equation}
where $\omega>0$ and $a$ is a constant given by
\begin{equation}
a = \frac{1}{2}\big[\beta_i+(2/b)\big]^{1/2}\bigg( \frac{\pi}{\beta_i} 
\frac{T_e}{T_i} \frac{m_e}{m_i}\bigg)^{1/2} \left\{1-\frac{1+b\beta_i}
{2[1+(b\beta_i/2)]^2}\right\}.
\label{nu}
\end{equation}
Here, $m_e$ is the electron mass, $m_i$ is the proton mass, $T_e$ is the 
equilibrium electron temperature, $T_i$ is the equilibrium proton temperature, 
$b=1+T_e/T_i$, and $\beta_i=n_0k_BT_i/(B_0^2/2\mu_0)^2$, 
where $k_B$ is Boltzmann's constant, $n_0$ is the equilibrium particle number density,
$B_0$is the equilibrium magnetic field, and $\mu_0$ is the permeability of free space 
(SI units).  Equation (\ref{ratio}) is valid when $k_\perp \rho_i \gg 1$ and 
$k_\perp \rho_e \ll 1$.  Using typical plasma parameters for the solar wind at 1 AU, 
equation (\ref{ratio}) was compared to the ratio $\gamma/\omega$ obtained from the
hot plasma dispersion relation and found to be an excellent approximation except
when $k_\perp \rho_i \sim 1$ where (\ref{ratio}) underestimated the damping although it
remained accurate to within a factor of 2 or so.  
%\smallskip

The substitution of equation (\ref{ratio}) into (\ref{CE2}) yields the equation
\begin{equation}
\frac{d(\log \eps)}{dk} = -\frac{4\pi a}{A}\exp(4\pi a k),
\label{CE3}
\end{equation}
where $k=k_\perp \rho_i$.  The solution is given by
\begin{equation}
\frac{\eps}{\eps_0} = \exp\left[ -\frac{1}{A}\big(e^{4\pi a k}- e^{4\pi a k_0}\big)\right].
\label{soln1}
\end{equation}
If the exponential factor is omitted from equation (\ref{CE2log}) or, equivalently, 
equation (\ref{KR}) is used in place of (\ref{KR2}), then instead of equation (\ref{CE3}) one obtains
\begin{equation}
\frac{d(\log \eps)}{dk} = -\frac{4\pi a}{A}
\label{CE4}
\end{equation}
which has the solution
\begin{equation}
\frac{\eps}{\eps_0} = \exp\left[ -\frac{4\pi a}{A}(k-k_0)\right].
\label{soln2}
\end{equation}
Equations (\ref{soln1}) and (\ref{soln2}) describe how the energy cascade rate changes as a result 
of collisionless damping.  The associated energy spectrum can be obtained from equation 
(\ref{reln2}) if the propagation angle $\theta(k)$ is known since the propagation angle is 
necessary to compute the wave frequency $\omega(k)$.  Therefore, the energy spectrum can only be 
obtained by solving the complete model consisting of equations 
(\ref{reln2}), (\ref{CE2log}), (\ref{CB}), and (\ref{CB1}).  This is not attempted here because
the energy spectrum is not needed for the purpose of this study.
%\smallskip

The theoretical model developed here may also be adapted to study hydrodynamic turbulence.
In hydrodynamics, the damping rate is $\gamma=-\nu k^2$, where $\nu$ is the kinematic viscosity,
and the cascade time is the nonlinear eddy turnover time $\tau = 1/kv$, where $v=(kE)^{1/2}$.
The model is based on equation (\ref{CE}) and either (\ref{KR}) or (\ref{KR2})
with $\alpha=\exp(-\nu k^2 \tau)$.  An analytic solution is only possible when 
(\ref{KR}) is used.  Calculations of the energy cascade rate versus wavenumber obtained using 
these hydrodynamic models show that the energy cascade rate appoaches zero near the 
Kolmogorov scale $\eta \sim (\nu^3/\eps)^{1/4}$.
Moreover, the model calculations in the range $k\eta \lesssim 1$ are in reasonable agreement 
with the cascade rate $\eps(k)$ obtained using the model spectrum in \citet{Pope:2000}.  
In fact, the wavenumber $k\eta \simeq 1$ where the energy cascade terminates in Pope's solution 
lies between the cutoff wavenumbers $k\eta \simeq 0.8$ and $k\eta \simeq 1.3$ for the two 
solutions obtained using either (\ref{KR}) or (\ref{KR2}).

\section{Results}

%\indent\indent
Results are now presented for typical high-speed solar wind in the ecliptic
%-----------------------------------------------------------------------------------------------
\begin{figure}
\begin{center}
\includegraphics[width=2.5in]{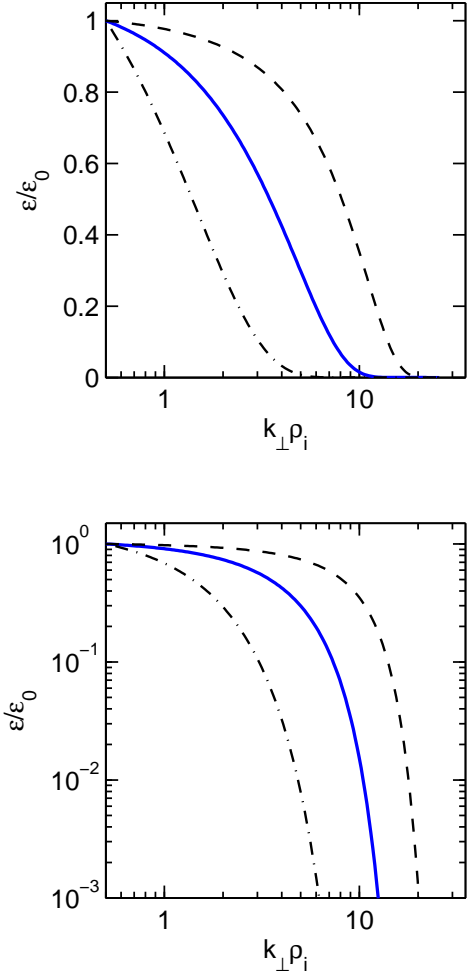}% 
\caption{\label{fig3}%
     The energy cascade rate $\epsilon(k)$ of the KAW cascade in the dissipation range
obtained from equation (\ref{soln1}) for conditions typical of high-speed solar 
wind at 1 AU: $\beta_i=1$, $\beta_e=1/2$, and $v_{\rm th, i}/c=2\times 10^{-4}$. 
Solutions are shown for $A=1$ (blue), $A=4$ (dashed), and $A=1/4$ (dot-dashed).
The same plot is displayed using a linear scale for the vertical axis in the upper panel 
and a logarithmic scale in the lower panel.
}
\end{center}
\end{figure}
%-----------------------------------------------------------------------------------------------
%-----------------------------------------------------------------------------------------------
\begin{figure}
\begin{center}
\includegraphics[width=2.5in]{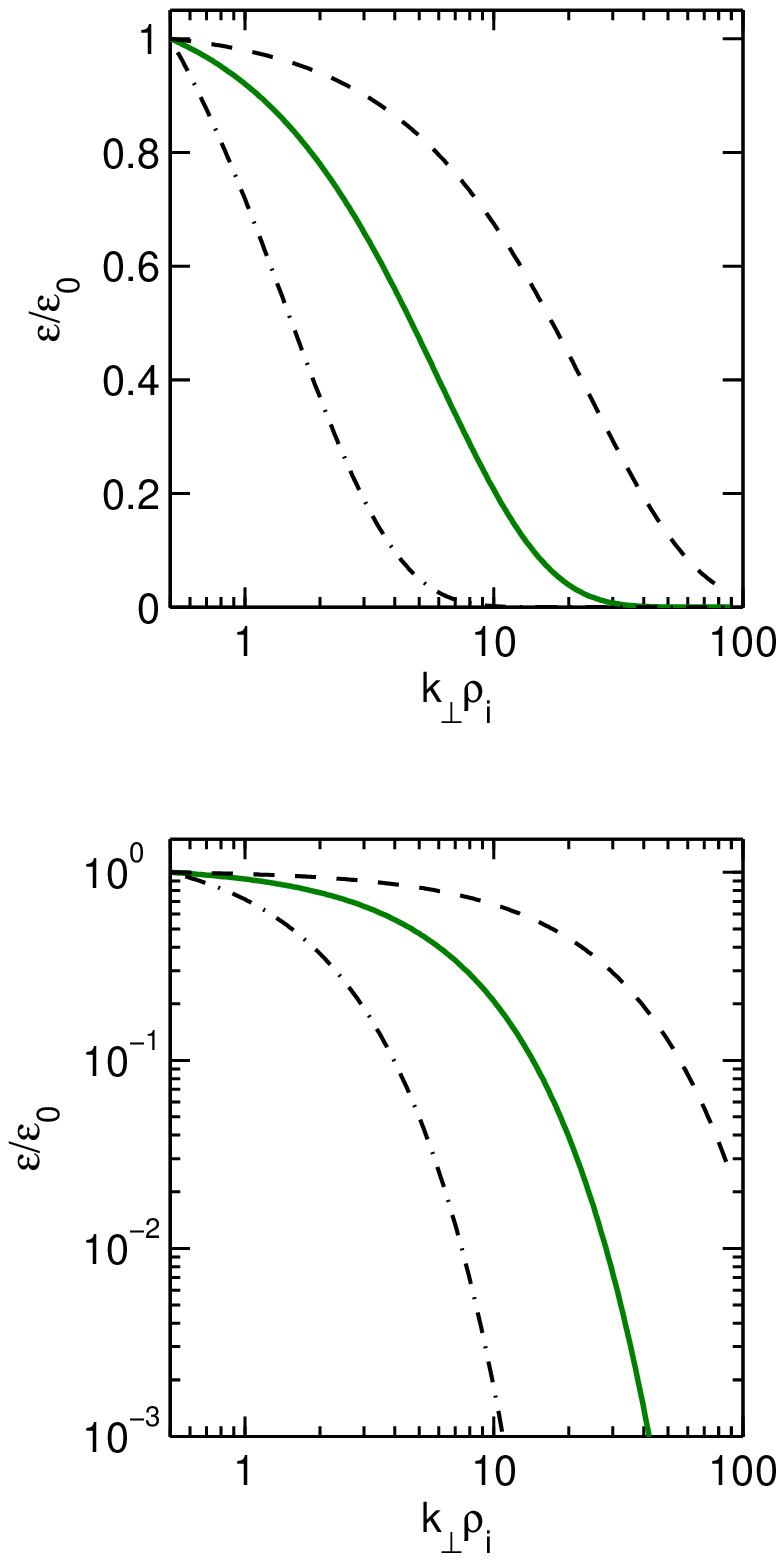}% 
\caption{\label{fig4}%
     The energy cascade rate $\epsilon(k)$ of the KAW cascade in the dissipation range
obtained from equation (\ref{soln2}) for conditions typical of high-speed solar 
wind at 1 AU: $\beta_i=1$, $\beta_e=1/2$, and $v_{\rm th, i}/c=2\times 10^{-4}$. 
Solutions are shown for $A=1$ (green), $A=4$ (dashed), and $A=1/4$ (dot-dashed).
The same plot is displayed using a linear scale for the vertical axis in the upper panel 
and a logarithmic scale in the lower panel.
}
\end{center}
\end{figure}
%-----------------------------------------------------------------------------------------------
plane near 1 AU with the plasma parameters $V_{\rm sw}>600$ km/s, 
$\beta_i=1$, $\beta_e=1/2$, and $v_{\rm th, p}/c=2\times 10^{-4}$.  This choice
of parameters is based on some of our own work, both published \citep{Podesta:2009b} and 
unpublished, and results published in the literature \citep{Feldman:1977,Newbury:1998,Schwenn:2006}.
For these parameters and for highly oblique angles of propagation  
($80^\circ < \theta < 90^\circ$), the damping of KAWs
is negligible for small wavenumbers, $k_\perp \rho_i \ll 1$.  KAW damping starts
to become significant around $k_\perp \rho_i \simeq 1/2$.   
The initial wavenumber in the model calculations is $k_\perp \rho_i \simeq 1/2$.  
%Measurements of power spectra of magnetic field fluctuations in the solar wind
%indicate that at the highest-wavenumbers in the inertial range, that is, at 
%spacecraft-frame frequencies near 0.1 Hz, it is typical to find $\delta v/v_A\simeq 1/10$.
%Therefore, it follows from the assumption of critical balance that 
%$k_\perp/k_\parallel \simeq 10$.  This yields a propagation angle of 84.3 degrees.
The results for the two different theoretical models, 
equations (\ref{soln1}) and (\ref{soln2}),
%%%%the initial condition $k_\perp/k_\parallel \simeq 10$ at $k_\perp \rho_i \simeq 1/2$ 
are shown in Figures \ref{fig3} and \ref{fig4}.
%\smallskip

The energy cascade rate $\eps$ shown in Figures~\ref{fig3} and \ref{fig4}
decreases to zero around $k_\perp \rho_i \simeq 10$ and $k_\perp \rho_i \simeq 25$,
respectively.  When $\eps/\eps_0 \ll 1$ the energy cascade terminates.  
Even though the two models yield noticeably different results, the wavenumber
where the cascade rate becomes negligibly small, $\eps/\eps_0 \ll 1$,
is of the same order of magnitude in both Figures \ref{fig3} and \ref{fig4}.
Thus, the two models are roughly consistent with each other and show that the KAW energy 
cascade cannot continue beyond the wavenumber 
$k_\perp \rho_i \sim 25$ because the energy flux is almost completely damped 
at that point.  This is the central conclusion of this study.  
The behavior of the solutions when the parameter $A$ is varied are also shown in 
Figures \ref{fig3} and \ref{fig4}.  The model (\ref{soln1}) that includes
the factor $\alpha^2$ in Kolmogorov's relation (\ref{KR2}) is less sensitive
to variations in the parameter $A$ than the model (\ref{soln2}) that is
based on Kolmogorov's relation (\ref{KR}).  Taken together, the results in 
Figures \ref{fig3} and \ref{fig4} suggest that the KAW cascade will be completely 
damped by collisionless Landau damping before the energy can reach the scale 
of the electron gyro-radius; the electron gyro-radius occurs at $k_\perp \rho_i \simeq 60$ 
for the plasma parameters used here.
%\smallskip

The wavenumber where the KAW cascade terminates can be estimated from
equations (\ref{soln1}) and (\ref{soln2}).  The wavenumber where $\eps/\eps_0$
reaches some small value, say $\eps/\eps_0=\delta$ with $\delta = 10^{-2}$,
can be determined as follows.  From equation (\ref{soln1}), assuming $4\pi k_0 a\ll 1$,
the wavenumber where $\eps/\eps_0=\delta$ is given by
\begin{equation}
k_\perp \rho_i \simeq \frac{1}{4\pi a} \log[1+A\log(1/\delta)],
\label{cutoff1}
\end{equation}
where $a$ is defined in equation (\ref{nu}).  
For $A=1$ and $\delta = 10^{-2}$, this yields $k_\perp \rho_i \simeq 10$.  From the 
second solution (\ref{soln2}), the wavenumber where $\eps/\eps_0=\delta$ is given by
\begin{equation}
k_\perp \rho_i \simeq \frac{A}{4\pi a} \log(1/\delta).
\label{cutoff2}
\end{equation}
For $A=1$ and $\delta = 10^{-2}$, this yields $k_\perp \rho_i \simeq 28$.  These 
formulas are very convenient for estimating the wavenumber where the KAW cascade terminates.

The termination point can also be expressed in terms of the
ratio $\gamma /\omega$.  Using equation (\ref{ratio}), the condition (\ref{cutoff1})
with $A=1$ and $\delta=10^{-2}$ is equivalent to $|\gamma /\omega|=0.14$.
Similarly, the condition (\ref{cutoff2}) with $A=1$ and $\delta=10^{-2}$ is equivalent 
to $|\gamma /\omega|=0.37$.  The application of theoretical 
models like those in Section 3 to hydrodynamic turbulence suggests that the actual 
termination point is somewhere between these two model predictions, $|\gamma /\omega|=0.14$ 
and $|\gamma /\omega|=0.37$.  Hence, it is reasonable to conclude that the cutoff for the 
KAW cascade occurs approximately 
when $|\gamma /\omega| \sim 0.25$.  Note that for the heuristic model in section 2
the cascade terminates as soon as $\alpha^2=\exp(4\pi \gamma/\omega)\ll 1$ which implies 
$|\gamma /\omega| \sim 0.18$.
Thus, the cutoffs for the KAW cascade predicted by the heuristic model in 
section 2 and the more elaborate models presented in section 3 agree to within a 
factor of 2.

\section{Conclusions}

%\indent
\indent
In this study, two different methods were employed to calculate the collisionless damping 
of the KAW cascade.  The first method demonstrates the effect of compounding
on wave dissipation during the energy cascade process as 
illustrated by the calculation in Table-1. For nearly perpendicular propagating 
KAWs under typical solar wind conditions at 1 AU, the ratio $|\gamma/\omega|$ 
is usually small from $k_\perp \rho_i= 1$ to $k_\perp \rho_i= 10$ so that 
$|\gamma/\omega|\ll 1$. 
Nevertheless, because the wave damping is compounded at each stage
of the energy cascade process the energy cascade 
is damped more rapidly than might be expected from an examination of the wavenumber
dependence of $\gamma/\omega$.  The simple heuristic model used to quantify this 
compounding effect assumes that all 
the energy is carried by KAWs, that the cascade time is equal to the wave period, and that
damping of these waves is governed by the linear Vlasov-Maxwell dispersion relation.  
The second method based on equation (\ref{CE2log}) yields quantitatively similar
results eventhough the precise form of Kolmogorov's relation is unknown.  
Thus, both methods support the main 
conclusion of this work, that a cascade consisting solely of KAWs cannot reach the 
electron gyro-scale in the solar wind at 1 AU.  This conclusion is also supported by 
the somewhat different cascade model developed by \citet{HowesJGR:2008a}.
%\smallskip

The conclusions are, of course, subject to some uncertainty.  For example,
if the cascade time is reduced from one wave period to half a wave period, then
the wavenumber where the KAW cascade terminates is roughly twice as large.
And, if the cascade time is reduced to one quarter of a wave period, then
the wavenumber where the KAW cascade terminates is roughly four times as large.
Consequently, the results are sensitive to the cascade time which is not known
precisely.

A simple expression for the wavenumber where the KAW cascade terminates has been obtained from 
the analytical solutions (\ref{soln1}) and (\ref{soln2}).  For $A=1$, the cutoff occurs 
approximately when $|\gamma/\omega|\simeq 0.25$.  Hence, one must be careful when using 
damping rates obtained from the linear Vlasov-Maxwell wave theory to estimate the stopping 
point of the cascade.  As a consequence of the effects of compounding, the wavenumber  
where the cascade terminates occurs not when $|\gamma/\omega|\simeq 1$ but
approximately when $|\gamma/\omega|\simeq 0.25$.
%%\smallskip

The conclusion reached in the study of \citet{Sahraoui:2009} should be reexamined in 
light of this result.  \citet{Sahraoui:2009}  have suggested that high-frequency 
magnetic field spectra in the solar wind may be caused by a KAW cascade from proton 
to electron scales.  To support this idea \citet{Sahraoui:2009} compute dispersion curves
from the hot plasma dispersion relation using
the solar wind parameters in their measurements and show that the ratio $|\gamma/\omega|$
increases to values of order $|\gamma/\omega|\simeq 1$, indicating strong damping, when
the wavenumber reaches the scale of the thermal electron gyro-radius $k\rho_e=1$.
The physical interpretation suggested by \citet{Sahraoui:2009} appears to be untenable 
based on the model calculations
presented here.  Using the dispersion curves in Figure 5 of the paper
by \citet{Sahraoui:2009}, the condition for the termination of the KAW cascade
$|\gamma/\omega|\simeq 0.25$ occurs when $k\rho_p\simeq 30$.  Hence, the KAW cascade 
cannot reach the electron gyro-scale at $k\rho_p\simeq 95$ and cannot by itself account 
for the high frequency solar wind spectra reported by \citet{Sahraoui:2009}.  This means 
that at least part of the high-frequency power-law spectrum measured in the solar wind by 
\citet{Sahraoui:2009}, \citet{Kiyani:2009}, and \citet{Alexandrova:2009} must be 
supported by some other kind of wave modes.  
%\smallskip

The theoretical model developed here depends crucially on the wave damping rates
computed from the linear dispersion relation and on the hypothesis of critical balance.
Particle distribution functions in the solar wind are usually
anisotropic with high energy tails and sometimes other features which can deviate 
significantly from the isotropic Maxwell distributions considered here. The effects 
of these features on the wave damping rates need to be taken into account in a more 
thorough analysis. If the hypothesis of critical balance fails to hold or is modified
somehow in the dissipation range, then the theory presented 
here would be incorrect and would have to be revised.  A separate effect that needs to be
taken into account for the interpretation of solar wind measurements is the breakdown of 
Taylor's hypothesis caused by the increasing 
phase velocities of the waves for $k_\perp\rho_i\gtrsim 1$.  This should reduce
the magnitude of the spectral exponent somewhat by systematically shifting power to 
higher wavenumbers, thereby creating a shallower spectrum.  
%\smallskip

%-----------------------------------------------------------------------------------------------
\begin{figure}
\begin{center}
\includegraphics[width=2.5in]{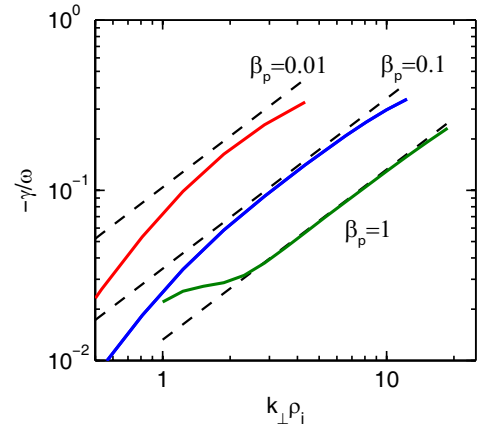}% 
\caption{\label{fig5}%
     The ratio $\gamma/\omega$ obtained from the hot plasma dispersion relation for
a propagation angle $\theta=89.9$ degrees
is compared to the approximation (\ref{ratio}) used in the theoretical model (dashed)
for $\beta_p=\beta_e=10^{-2}$ and $T_p=10^6$ K (red),
$\beta_p=\beta_e=10^{-1}$ and $T_p=10^6$ K (blue), and 
$\beta_p=1$, $\beta_e=1/2$ and $v_{\rm th, p}=60$ km/s (green)}
\end{center}
\end{figure}
%-----------------------------------------------------------------------------------------------
It is of interest to apply the model developed here to study
the damping of MHD turbulence in the solar corona.  Because the plasma
parameters vary significantly between the low and high corona and the particle
distribution functions are nonthermal, a survey of parameter space 
is required.  This is an important avenue of investigation for future work.  It is also of
some practical importance since NASA plans to launch a spacecraft into the high corona 
in the coming decade \citep{McComas:2007}.  The Solar Probe will be the first spacecraft
to reach perihelion in the ecliptic plane at a heliocentric distance of
approximately 10 solar radii where the plasma beta is around $1/10$.

For the representative coronal parameters $\beta_p=\beta_e=0.1$ and $T_p=10^6$ K
the prediction for an isothermal electron-proton plasma is that the KAW cascade 
terminates when $k_\perp\rho_i\sim 7$.  For $\beta_p=\beta_e=10^{-2}$ and $T_p=10^6$ K
the theory predicts that the KAW cascade terminates when $k_\perp\rho_i\sim 2.5$.
It should be noted that for these plasma parameters the approximation (\ref{ratio})
overestimates the damping rate at $k_\perp \rho_i\sim 1$ by a factor of $\sim 1.5$
and, consequently, these preliminary estimates could be too small by a factor of 2 or so.
Nevertheless, these preliminary calculations suggest that
the KAW cascade cannot reach the electron gyro-scale in the solar corona.

The cascade terminates more rapidly at low $\beta_p$
because for a fixed value of $k_\perp \rho_i$ the ratio $|\gamma/\omega|$ 
increases as $\beta_p$ decreases.  This is shown in Figure-\ref{fig5} where the 
ratio $\gamma/\omega$ obtained from the hot plasma dispersion relation is compared 
to the approximation (\ref{ratio}) used in the theoretical model.  Note from 
equations (\ref{ratio}) and (\ref{nu}) that when $T_e=T_p$ and $\beta_p\ll 1$, 
if $k_\perp \rho_i$ is held fixed ($k_\perp \rho_i\gg 1$),
then $|\gamma/\omega|\propto \beta_p^{-1/2}$ even though for a fixed value of 
$k_\perp c/\omega_{pi}$ the ratio $|\gamma/\omega|$ is independent of $\beta_p$.

%the approximate dispersion 
%relation $\omega \simeq k_\parallel v_A \sqrt{1+(k_\perp \rho_i)^2}$ that 
%the electron resonance condition $\omega/k_\parallel v_e \lesssim 1$ is satisfied when 
%\begin{equation}
%\sqrt{1+(k_\perp \rho_i)^2} \lesssim \bigg(\frac{m_p}{m_e}\frac{\beta_p}{2}\bigg)^{1/2}
%%\label{CE4}
%\end{equation}
%which implies
%\begin{equation}
%k_\perp \rho_i \lesssim \left\{ \begin{array}{ll} 9.6 & \mbox{if $\beta_p=10^{-1}$}, \\
%3.0 & \mbox{if $\beta_p=10^{-2}$}. \end{array} \right.
%%\label{CE4}
%\end{equation}
%This behavior appears contrary to that of long wavelength Alfv\'en waves for which   
%electron Landau damping is an increasing function of $\beta$ \citep{GaryBorovsky:2008}.
%However, when plotted as a function of $k_\perp c/\omega_{pi}$, 
%the ratio $|\gamma/\omega|$ increases as $\beta_p$ increases.  This is a well
%known feature of long wavelength Alfv\'en waves, that is,
%electron Landau damping is an increasing function of $\beta$ \citep{GaryBorovsky:2008}.

This work was supported by the NASA Solar and Heliospheric Physics Program,
the NASA Heliospheric Guest-Investigator Program, the 
NSF SHINE Program, and by the Los Alamos National Laboratory LDRD Program.

\bibliographystyle{apj}
\bibliography{jp}

\IfFileExists{\jobname.bbl}{}
 {\typeout{}
  \typeout{******************************************}
  \typeout{** Please run "bibtex \jobname" to optain}
  \typeout{** the bibliography and then re-run LaTeX}
  \typeout{** twice to fix the references!}
  \typeout{******************************************}
  \typeout{}
 }

\end{document}